\documentstyle[psfig,11pt,a4]{article}

\newcommand{\be}{\begin{equation}}
\newcommand{\ee}{\end{equation}} 
\newcommand{\bea}{\begin{eqnarray}}
\newcommand{\eea}{\end{eqnarray}}

\begin{document}

\begin{titlepage}

\begin{flushright} 
{\tt 	FTUV/99-14\\
	IFIC/99-14\\
	SU-ITP/99-10\\
	gr-qc/9902084}
 \end{flushright}

\bigskip

\begin{center}

{\Large {\bf On the normalization of Killing vectors and energy\\
conservation in two-dimensional gravity
\footnote{Work partially supported by the 
{\it Comisi\'on Interministerial de Ciencia y Tecnolog\'{\i}a}\/ 
and {\it DGICYT}.}}
}

\bigskip 
\bigskip\bigskip
 J. Cruz$^a$ \footnote{\sc cruz@lie.uv.es}, A. 
 Fabbri$^b$ \footnote{ \sc afabbri1@leland.stanford.edu} and
 J. Navarro-Salas$^a$ \footnote{\sc jnavarro@lie.uv.es},

\end{center}

\bigskip%

\footnotesize
\noindent	
 a) Departamento de F\'{\i}sica Te\'orica and 
	IFIC, Centro Mixto Universidad de Valencia-CSIC.
	Facultad de F\'{\i}sica, Universidad de Valencia,	
        Burjassot-46100, Valencia, Spain. 
 \newline                 
b) Department of Physics, Stanford University, Stanford, CA, 94305-4060, USA.
\normalsize 

\bigskip

\bigskip

\begin{center}
{\bf Abstract}
\end{center}
We explicitly show that, in the context of a recently proposed 2D dilaton
gravity theory, energy conservation requires the ``natural'' Killing vector
to have, asymptotically, an unusual normalization. The Hawking temperature
$T_H$ is then calculated according to this prescription.
 

\end{titlepage}

\newpage
Two-dimensional dilaton gravity theories are considered interesting
toy models for studying the issues connected with quantization of gravity.
In particular,  the use of exactly solvable models 
makes it possible to follow, analytically,  
the semiclassical evolution of evaporating black holes.  
In the context of the one discovered by  
RST \cite{RST}, 
semiclassical version of the CGHS theory \cite{CGHS}, black
holes evaporate completely and the final end-point geometry is the
vacuum state. It makes sense, then, to check whether the same
conclusions hold in more general theories. Quite interestingly, it was
shown recently \cite{rescaled} that two models related by a conformal
rescaling of the
metric may indeed describe very different physics: in one model, the so
called exponential model introduced in \cite{exp}, the evaporation process
never ceases, whereas in the rescaled theory black holes
 evaporate completely leaving an everywhere regular end-point geometry 
that may be regarded as the semiclassical ground state (as it happens,
for instance, in \cite{BPP}). Several interesting questions are
raised from this analysis,  both in the classical and the
semiclassical theory. In this note we will restrict to the classical
analysis of black hole formation in the rescaled theory \cite{rescaled} 
and show that energy conservation
 implies a nontrivial normalization for the
asymptotic timelike
Killing vector
associated to ``natural'' observers at rest at infinity. \\
Let us start by reminding briefly what happens in the CGHS
theory \cite{CGHS},
described by the action
\be
S={1\over2\pi}\int d^2x\sqrt{-g}\left[e^{-2\phi}\left( R+ 4(\nabla\phi)^2 
+4\lambda^2\right) -{1\over2}
\sum_{\i=1}^N(\nabla f_i)^2\right]
\>.\label{cghs}
\ee
In the Kruskal gauge the static black hole solutions are
\be
ds^2= \frac{-dx^+dx^-}{{M\over \lambda}-\lambda^2x^+x^-}, \ \ 
e^{-2\phi}={M\over\lambda} -\lambda^2 x^+x^- \>,\label{soli}
\ee
where $M$ is identified with the ADM mass. A black hole can be formed from
the vacuum by sending in a shock-wave at $x^+=x_0^+$, with
$T_{++}={M\over {\lambda x_0^+}}\delta (x^+-x_0^+)$.
The line element, that for $x^+<x_0^+$ was given by
$ds^2=\frac{dx^+dx^-}{\lambda^2 x^+x^-}$, in the future of the shock-wave 
becomes 
\be
ds^2 =-\frac{dx^+dx^-}{ {M\over\lambda} -\lambda^2 x^+(x^-
+{M\over{\lambda^3 x_0^+}})} \>.\label{bho}
\ee
Provided we introduce the asymptotically minkowskian coordinate 
$\lambda x^+=e^{\lambda\sigma^+}$ energy
conservation is then simply given by the fact that 
\be
E=\int d\sigma^+ T_{\sigma^+\sigma^+} = \lambda \int dx^+ x^+ T_{++}
\label{ene}
\ee
equals the mass $M$ of the black hole formed. \\ 
The rescaled exponential model of \cite{rescaled}, namely
\bea
S&=&{1\over2\pi}\int d^2x\sqrt{-g} \left[R\phi+{\beta e^{\beta\phi}\over
e^{\beta\phi}-1}(\nabla\phi)^2\right.\nonumber\\&&\left.
+{4\lambda^2\over\beta}e^{\beta\phi}\left(e^{\beta\phi}-1\right)-{1\over2}
\sum_{\i=1}^N(\nabla f_i)^2\right]
\>,\label{expres}
\eea
has the following black hole solutions in Kruskal gauge
\be
ds^2=\frac{-dx^+dx^-}{{1\over\beta}-{\lambda^2\over C}-{C\over\beta}x^+x^-}
\>, \label{i}
\ee
\be
e^{-\beta\phi}=
{{\lambda^2\beta}\over C} + C x^+x^- \>.\label{dila}
\ee
It is useful, in this context, to consider a process a bit more
general than the one described earlier. 
Let us study what happens when we send  a shock-wave to a 
black hole  characterized by $C=C_{in}$ (the case
$C_{in}=\lambda^2\beta$, corresponding to the Minkowski ground state, was
considered in \cite{rescaled}) .
The "out" solution will be given by\be
ds^2=\frac{-dx^+dx^-}{{1\over\beta}-{\lambda^2\over C_{out}}-{C_{out}
\over\beta}(x^++\Delta^+)(x^-+\Delta^-)}
\>,\label{ops}
\ee
and by imposing continuity to the metric across the shock-wave line 
$x^+=x^+_0$,
we can determine the value of $\Delta^{\pm}$
\be
\Delta^+=x^+_0\left({C_{in}\over C_{out}}-1\right)
\>,\ee
\be
\Delta^-={\lambda^2\beta\over x_0^+ C_{in}}\left({1\over C_{in}}-{1\over C_{out}
}\right)
\>.\ee
We note, by comparison to the CGHS theory, that an additional shift has
appeared, namely $\Delta^+$. 
The energy momentum tensor has the expression
\be
T^f_{++}={\Delta^-C_{in}C_{out}\over \lambda^2\beta^2}\delta (x^+-x^+_0)=
{C_{out}\over\beta x_0^+}\left({1\over C_{in}}-{1\over C_{out}}\right)
\delta (x^+-x^+_0)\>.
\ee
In order to calculate the energy of this shock-wave
 we must write down $T^f_{++}$
in asymptotically minkowskian coordinates.
However, due the presence of a nonvanishing $\Delta^+$  an ambiguity 
arises.
The asymptotically minkowskian coordinate $\sigma^+$
is different for the in and out solutions, namely
\be
\sqrt{{C_{in}\over\beta}}x^+=e^{\sqrt{{C_{in}\over\beta}}\sigma^+},\qquad
\sqrt{{C_{out}\over\beta}}(x^++\Delta^+)=e^{\sqrt{{C_{out}\over\beta}}\tilde
\sigma^+}\>,
\ee
and the energy momentum tensor in terms of these coordinates becomes
\be
T^f_{\sigma^+\sigma^+}={1\over\beta}\sqrt{{C_{in}\over\beta}}\left(
{C_{out}\over C_{in}}-1\right)
\delta (\sigma^+-\sigma^+_0)
\>,\ee
\be
T^f_{\tilde\sigma^+\tilde\sigma^+}={1\over\beta}\sqrt{{C_{out}\over\beta}}\left(
1-{C_{in}\over C_{out}}\right)
\delta (\tilde\sigma^+-\tilde\sigma^+_0)
\>,\ee
giving, potentially, two different results for the energy 
of the shock wave.
Furthermore, neither of these results can be 
compatible with energy conservation because 
they do not satisfy the basic requirement
\be
E=f(C_{out})-f(C_{in})\>,\label{f}
\ee
for some function $f$, which expresses that the energy of the
 shock wave must be equal to
the difference between the ADM masses of the in and out solutions.
We can avoid this problem if we just state that the energy of the 
shock-wave is measured by the observer
 in which the metric asymptotically behaves as
 \be
 ds^2=-{C\over\lambda^2\beta}d\sigma^+d \sigma^-
 \>. \label{nuco}
\ee
 These new coordinates are defined by
 \be
 {C_{in}\over{\lambda\beta}}
x^+=e^{{C_{in}\over{\lambda\beta}}\sigma^+},\qquad 
 {C_{out}\over{\lambda\beta}} (x^+
+\Delta^+)=e^{{C_{out}\over{\lambda\beta}}\tilde \sigma^+}\>,\ee
and now the energy momentum tensor becomes
\be
T^f_{\sigma^+\sigma^+}={1\over\lambda\beta^2}(C_{out}-C_{in})
\delta (\sigma^+-\sigma^+_0)
\>,\label{xii}\ee
\be
T^f_{\tilde\sigma^+\tilde\sigma^+}={1\over\lambda\beta^2}(C_{out}-C_{in})
\delta (\tilde\sigma^+-\tilde\sigma^+_0)
\>.\label{xiii}\ee
This new result has the following two desired properties:
\begin{itemize}
\item The energy of the shock-wave does not depend on the choice of in or out
 observers (this can be easily understood by noting that we now have
$\frac{d\sigma^+}{d\tilde\sigma^+}=1$ at the shock-wave) ;
\item It satisfies the consistency condition (\ref{f}).
\end{itemize}
We must point out that this choice of coordinates implies that the
 normalization of
the time-like Killing vector $\xi=\partial_t$ at spatial infinity is
$-{C\over\lambda^2\beta}$ instead of the more standard one
 $\xi^2=-1$. 
 Now we must determine whether this result agrees with the energy
conservation.
 In doing so we need an expression for the ADM mass of the static solutions
 (\ref{i}), (\ref{dila}).
 Let us make use of the formula \cite{Mann} (see also \cite{gegemberg})
 \be
 M_{ADM}={F_0\over2}\left\{\int^{\phi}dD(s) V(s)
 e^{-\int^{s}dt {H(t)\over
 D^{\prime}(t)}}-(\nabla D(\phi))^2e^{-\int^{\phi} dt {H(t)\over D^{\prime}(t)}}\right\}
 \> \label{cucu}\ee
 for the ADM mass of a static solution arising from the action 
 \be
 S=\int d^2x \sqrt{-g}\left[D(\phi)R+H(\phi)(\nabla\phi)^2
 +V(\phi)\right]\>,\ee
 where in our case
 \be
 D(\phi)=\phi \>,\ee
 \be
 H(\phi)={\beta e^{\beta\phi}\over e^{\beta\phi}-1}\>,\ee
 \be
 V(\phi)={4\lambda^2\over\beta}e^{\beta\phi}(e^{\beta\phi}-1)
 \>.\ee
 Applying formula (\ref{cucu}) to the static solutions (\ref{i}),
 (\ref{dila}) we
 get
 \be 
 M_{ADM}= F_0\left({2C\over\beta^3}+constant\right) \>.\ee 
 The constant $F_0$
 is related to the normalization at
 infinity of the time-like Killing vector.
 Actually we have that  \be
 \lim_{x\rightarrow\infty} \xi^2=-\left({{2F_0}\over{\beta}}
 \sqrt{{C\over\beta}}\right)^2\>.\ee
 So if we take the same normalization we used when we calculated the energy of
 the shock-wave, we have $F_0={\beta\over2\lambda}$ and therefore
 \be
 M_{ADM}=
 {C\over{\lambda\beta^2}} +constant \>.\label{massa}
 \ee
 This way the energy of the shock-wave, given by the formulas (\ref{xii}),
 (\ref{xiii})
 is just
 \be
 E=M_{ADM}(out)-M_{ADM}(in)\>,\ee
 and the energy is exactly conserved.
 Thus we are led to conclude that the only possible way to achieve energy
 conservation in this model is to pick up a "mass dependent" normalization at
 infinity of the time-like Killing vector instead of the standard one. \\
 As a straightforward application of the above considerations, we now
 determine the Hawking temperature $T_H$ of the black holes (\ref{i}).  
 The usual argument of determining the period of Euclidean
 time $\tau=it$  in order for the
 metric to be free of conical singularities at the event horizon 
 gives
 \be
 T_H=\frac{C}{2\pi\lambda\beta}=\frac{1}{2\pi}(\lambda +\beta M)
 \>.\label{temp}
 \ee 
 In the last equality, according to \cite{rescaled}, 
 the constant appearing in (\ref{massa}) has been chosen in such a way
 that $M=0$ in the ground
 state $C=\lambda^2\beta$. 
 The key observation in obtaining this formula is that we used
 $t=\frac{\sigma^+ + \sigma^-}{2}$, where, again,  $\sigma^{\pm}$ are
 defined requiring
 the metric to be asymptotically of the form (\ref{nuco}).
 We note, finally, that the same result can be
 inferred by calculating the near-horizon evaporation flux in the 
 ``out'' region
    (\ref{ops}), i.e. 
 \be
 <T^f_{\tilde\sigma^-\tilde\sigma^-}>|_h=-{N\over24}\left\{\sigma^-,\tilde
 \sigma^-\right\}|_h={N\over48}{C^2\over\lambda^2\beta^2}={N\over12}\pi^2
 T_H^2
 \>,\ee
 where $\left\{\sigma^-,\tilde\sigma^-\right\}$ is the Schwartzian
 derivative and $\sigma^-$, $\tilde\sigma^-$ are defined by
  \be
 -{C_{in}\over{\lambda\beta}}
 x^-=e^{{-C_{in}\over{\lambda\beta}}\sigma^-},\qquad
 -{C_{out}\over{\lambda\beta}} (x^-
+\Delta^-)=e^{-{C_{out}\over{\lambda\beta}}\tilde \sigma^-}\>.\ee

The aim of this note has been to point out the unusual normalization required
for the Killing vector to produce a definition of the energy compatible
with its conservation in a physical process. 
\section*{Acknowledgements}
J. C acknowledges the Generalitat Valenciana for a FPI fellowship.
A. F. is supported by an INFN fellowship.

 \end{document}